# Low-Temperature Remote Plasma Synthesis of Highly Porous TiO$_2$ as Electron Transport Layers in Perovskite Solar Cells

*Jose M. Obrero-Perez, Fernando Nuñez-Galvez, Lidia Contreras-Bernal, Javier Castillo-Seoane, Gloria P. Moreno, Triana Czermak, Francisco J. Aparicio, Teresa C. Rojas, Francisco J. Ferrer, Ana Borras, Angel Barranco\*, and Juan R Sánchez-Valencia\**

Dr. J.M. Obrero-Perez, F. Nuñez-Galvez, Dr. L. Contreras-Bernal, Dr. J. Castillo-Seoane, G.P. Moreno, T. Czermak, Dr. F. J. Aparicio, Dr. T.C. Rojas, Dr. A. Borras, Dr. Angel Barranco, and Dr. J.R Sánchez-Valencia.
Nanotechnology on Surfaces and Plasma Laboratory, Materials Science Institute of Seville (CSIC-US), C/ Américo Vespucio 49, 41092, Seville, Spain.
E-mail: jrsanchez@icmse.csic.es, angelbar@icmse.csic.es

F. J. Ferrer
Centro Nacional de Aceleradores (CNA, CSIC-Universidad de Sevilla)

F. Nuñez-Galvez
Dpto. Física Aplicada I. Escuela Politécnica Superior Universidad de Sevilla. c/ Virgen de África 7, Seville E-41011, Spain

Dr. L. Contreras-Bernal
Dpto. Ingeniería y Ciencia de los Materiales y del Transporte. Escuela Politécnica Superior Universidad de Sevilla. c/ Virgen de África 7, Seville E-41011, Spain

Halide perovskite solar cells (PSCs) offer high efficiency and low costs, making them key for future photovoltaics. Optimizing charge transport layers is crucial, with porous TiO$_2$ widely used as electron transport layers (ETL) due to its energy alignment, transparency, and abundance. However, its efficiency relies on crystallinity requiring high-temperature processing (>450°C), increasing costs and limiting flexible substrates. Low-temperature wet-chemical methods face scalability issues due to material waste and hazardous solvents. In this context, plasma-based technologies are emerging as a more efficient and sustainable alternative to oxide-based ETLs.

This study presents the synthesis of TiO$_2$ layers using an advanced plasma method combining remote plasma-assisted vacuum deposition (RPAVD) and soft plasma etching (SPE) at mild temperatures (<200°C), allowing control of microstructure and porosity. The resulting nanocolumnar film, decorated with a highly porous aerogel-like layer, enhances optical and electronic properties. These plasma-synthesized TiO$_2$ layers are antireflective and improve the efficiency in porous n-i-p PSCs, matching the performance of high-temperature reference cells. These PSCs achieve a champion PCE of 14.6%, a high value compared to reference devices synthesized at 450°C. Impedance spectroscopy confirms high recombination resistance and stable capacitance, linked to improved perovskite crystallinity. Our results highlight the potential of the RPAVD+SPE approach for producing low-temperature efficient ETLs, providing a feasible, industrially scalable, and eco-friendly alternative for manufacturing flexible, high-performance photovoltaic devices.

## 1. Introduction

The growing demand for clean and renewable energy has driven research and development efforts toward more efficient and cost-effective photovoltaic technologies. In this context, perovskite solar cells (PSCs) have emerged as a promising alternative due to their high conversion efficiency, low manufacturing costs, and versatility in device engineering.[1–3] To maximize their performance, optimizing each solar cell component, especially the electron transport layers (ETLs), is crucial for efficient charge transport and extraction.[4] Many materials have been investigated as ETL in PSC applications; however, titanium dioxide (TiO$_2$) was the first and is still the leading actor in most of the relevant works in perovskite solar cells due to its appropriate conduction band level alignment with perovskite, high transmittance, chemical stability, environmental compatibility, mature fabrication process, and abundance.[5,6] In addition, porous TiO$_2$ has been utilized from the birth of the PSC technology to enhance the photovoltaic behavior by expanding the active-electrode surface area, facilitating a rapid diffusion of charge carriers, minimizing recombination, improving light absorption, and reducing reflection at the layer interface.[5,7] Several predominant phases of TiO$_2$, including anatase, rutile, and brookite, have been implemented in PSCs.[8,9] Anatase TiO$_2$ is the most common type due to its good charge transport properties. Traditional methods for synthesizing porous anatase TiO$_2$, such as spin-coating from nanoparticle dispersion, often demand high temperatures (>450 °C) to ensure proper crystallinity and sintering of the layer.[10] This step limits the compatibility with heat-sensitive substrates, where the maximum processing temperature for a flexible substrate is around 200 °C, for example, for polyethylene terephthalate (PET). Low-temperature manufacturing is also essential in two-terminal tandem PSCs since elevated temperatures can degrade the passivation layers in the bottom cell, thereby limiting the processing of the top cell.[11] Thus, developing PSCs at low temperatures offers several advantages: compatibility with flexible substrates and various perovskite crystallization methods and reducing manufacturing costs. Indeed, the pressing need to establish low-temperature methods for depositing porous $TiO_2$ to fabricate efficient and cost-effective PSCs has become a central point of PSC research geared toward practical applications. While methods such as spray-coating,[12] colloidal $TiO_2$,[13] and chemical deposition bath techniques promise to fabricate porous $TiO_2$ at low temperatures,[14] they often involve wet processes that use solvents with high levels of toxicity, presenting environmental and health risks, and difficult recyclability, limiting as well the scalability and posing cost-effectiveness challenges.

In recent years, plasma technology has emerged as a promising technology for synthesizing porous $TiO_2$ thin films at low temperatures.[15,16] The interest in this technology lies in the fact that it offers precise control over deposition parameters to produce thin films with tailored properties and is industrially scalable.[17] These properties make them especially attractive for ETL applications in the field of PSCs. Among these techniques, remote plasma-assisted vacuum deposition (RPAVD) is a new and innovative methodology for the low-temperature synthesis of tunable porous metal oxides. RPAVD induces physicochemical reactions between precursor molecules to develop polymer-like layers at significantly lower temperatures than conventional methods and without chemical restrictions.[18–21] This approach not only solves the limitations of high-temperature deposition techniques but also presents additional advantages, such as improved uniformity, fine control of the layers' properties, and the industrial scalability of the process. In the present work, RPAVD has been combined with soft plasma etching (SPE) at mild temperatures (below 200ºC) to obtain crystalline structures with adjustable porosity, which reach very high values above 85%. Furthermore, these films are highly transparent, with antireflective properties, i.e., their light transmittance is higher than the substrate in the visible and near-infrared range. Varying the synthesis gases, we successfully produced two types of $TiO_2$ with different porosity and morphology. This versatility enabled us to customize $TiO_2$ layers to meet specific requirements, enhancing their performance in perovskite solar cells. The optimized PSC architecture consists of a columnar layer with an aerogel-like top layer. In this structure, a $TiO_2$ layer naturally forms a continuous thin film at the FTO interface, preventing direct contact between the perovskite and the electrode. At the same time, the open spaces between the columns promote efficient perovskite infiltration while maintaining connectivity across the layer. Using this architecture, we fabricate a perovskite solar cell that achieves a notable champion efficiency comparable to that of perovskite cells with ETLs fabricated by conventional methods at high temperatures in our laboratory. This highlights our RPAVD method's effectiveness in synthesizing high-quality metal oxide porous films as a promising tool for developing highly efficient PSCs at lower temperatures.

**2. Experimental**

*2.1 Substrate cleaning.* Pilkington FTO substrates (TEC-15 from XOP Glass Company), with a sheet resistance of 12-14 $\Omega \cdot sq^{-1}$ and total transmittance of 84% in the visible, undergo a four-stage cleaning protocol. Initially, they are brushed using a Hellmanex solution in water (2% vol. Hellmanex) and rinsed with Milli-Q water. Next, they are subjected to ultrasonic bath washing for 15 minutes, following this solvent sequence: 1. Hellmanex solution (as used previously); 2. Milli-Q water; 3. Acetone; 4. Isopropanol. Following this process, substrates are dried using nitrogen flow. Substrates are treated with UV/O$_3$ for 15 minutes to complete the cleaning using a UV Ozone Cleaner from Ossila.

*2.2 Fabrication of columnar TiO$_2$ thin films via RPAVD-O$_2^*$.* Titanium(IV) phthalocyanine dichloride (TiPcCl, dye content of 95%) was procured from Sigma-Aldrich and utilized without further purification. FTO substrates were positioned within an ECR-MW plasma reactor operating at 2.45 GHz, between the plasma discharge and the Knudsen sublimation cell, with the substrates facing opposite to the glow discharge (sample to sublimation cell distance is 9 cm, and sample to glow discharge region is 6 cm). Before deposition, the system was evacuated to attain a base pressure of $10^{-6}$ mbar. TiPcCl molecules were sublimated under O$_2$ plasma conditions at 350 W. An oxygen flow of 6 sccm was dosed with a calibrated mass flow controller. The pressure was controlled via an automated butterfly valve regulator, which controls the pumping flow of the system to maintain the pressure at $10^{-3}$ mbar. The deposition rate and resulting thickness of the columnar TiO$_2$ were tracked using a quartz crystal microbalance (QCM), adjusting the temperature of the sublimation cell to maintain a growth rate of 0.4 Å·s$^{-1}$ (density of 0.5 g·cm$^{-3}$ in the QCM electronics). The substrates were at room temperature, which was unaffected by the plasma discharge. These samples were labeled as *Columnar*.

A batch of the *Columnar* samples underwent oxygen soft plasma etching (SPE) to induce complete oxidation and remove most of the carbon from the precursor (labeled as *Columnar+SPE*). For this purpose, samples were annealed to 200ºC and positioned facing the glow discharge at a distance of 6 cm. The plasma power was set at 400 W for 2 hours under a pressure of $10^{-3}$ mbar of oxygen.

*2.3 Fabrication of Columnar-Aerogel $C_1$ and $C_2$ layers by RPAVD-Ar*$^*$ *+ SPE.* In the same plasma reactor, maintaining an identical base pressure of $10^{-6}$ mbar, *Columnar* TiO$_2$ thin films, positioned under the same geometrical conditions between the plasma discharge and the Knudsen sublimation cell, were used to deposit on top of the RPAVD-Ar$^*$ layers. Here, the TiPcCl precursor underwent sublimation under Ar plasma at 150 W, while the gas pressure was regulated at $10^{-2}$ mbar (introducing 25 sccm of Ar). The growth rate was sustained at 0.5 Å·s$^{-1}$ (density at 0.5 g·cm$^{-3}$ in QCM electronics) while the temperature remained at room temperature, as previously described. A plasma polymer of TiPcCl was formed on the columnar thin films. Subsequently, samples were subjected to oxygen-soft plasma etching following the abovementioned procedure. Cycles of plasma polymer deposition and plasma etching were conducted by rotating the sample holder and using the same experimental conditions as in the previous case to fabricate one (*Columnar-Aerogel $C_1$*) and two (*Columnar-Aerogel $C_2$*) cycles.

The descriptive sequence of all RPAVD TiO$_2$ samples is shown in Scheme 1. Four different samples were analyzed in this work, i.e.: *Columnar*, *Columnar+SPE*, *Columnar-Aerogel $C_1$,* and *Columnar-Aerogel $C_2$,* which correspond to *(RPAVD-O$_2$$^*$), (RPAVD-O$_2$$^*$+SPE), (RPAVD-O$_2$$^*$+SPE+RPAVD-Ar$^*$+SPE),* and *(RPAVD-O$_2$$^*$+SPE+RPAVD-Ar$^*$+SPE+RPAVD-Ar$^*$+SPE)*, respectively.

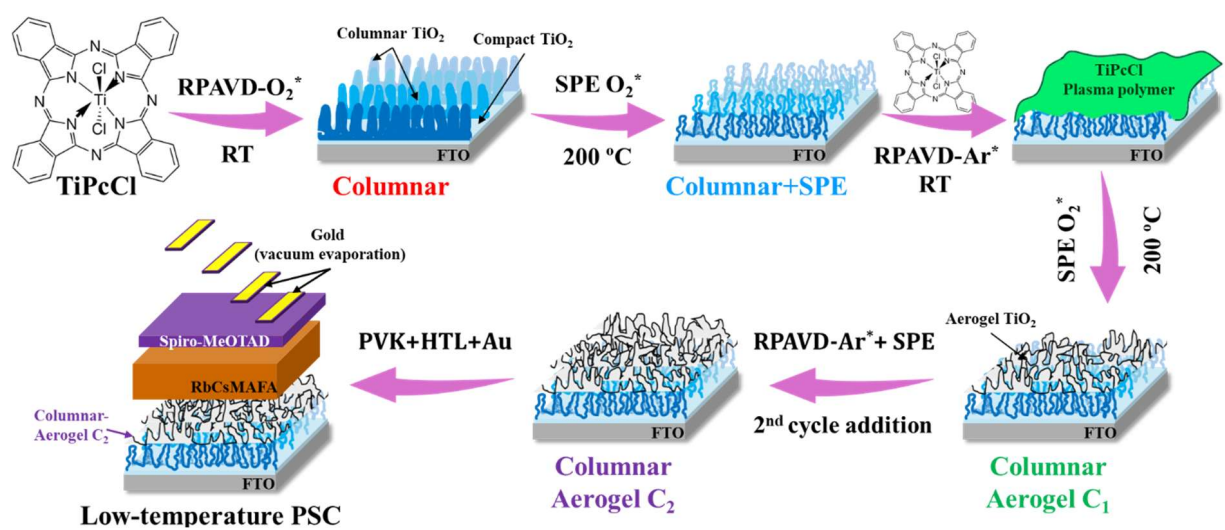

**Scheme 1.** Schematic representation of the fabrication process and descriptive sequence of all TiO$_2$ samples. This includes the fabrication of *Columnar* TiO$_2$ thin films via RPAVD-O$_2$$^*$, *Columnar+SPE,* and the subsequent formation of *Columnar-Aerogel C1* and *C2* layers using RPAVD-Ar$^*$ followed by SPE.



*2.4 Fabrication of compact and mesoporous TiO$_2$ layers for PSC reference.* Before manufacturing the mesoporous TiO$_2$ layer, a compact TiO$_2$ (c-TiO$_2$) layer must be deposited onto the clean FTO substrate. This compact layer acts as a barrier, preventing direct contact between the perovskite and the FTO substrate. The deposition of the c-TiO$_2$ film is accomplished using the spray-pyrolysis technique. A solution comprising 1 mL of titanium diisopropoxide bis(acetylacetonate) in 14 mL of absolute ethanol is prepared. This titanium solution is then uniformly spread onto the FTO substrates annealed to 450ºC, covering an approximate 100-200 cm$^2$ area, with oxygen as the carrier gas. After exhausting the compact TiO$_2$ solution, substrates are maintained at 450 °C for 30 minutes to ensure the formation of the anatase phase, achieving an approximate thickness of 20 nm. Subsequently, the substrates are allowed to cool to room temperature. Finally, the compact film undergoes UV/O$_3$ treatment for 15 minutes before the deposition of the mesoporous layer. It is worth noting that this c-TiO$_2$ has not been deposited for devices incorporating RPAVD layers to maintain the synthesis temperature below 200ºC.

The preparation of the subsequent mesoporous TiO$_2$ begins by diluting 150 mg of commercial TiO$_2$ paste in 1 mL of absolute ethanol. The solution is stirred 24 hours before use to ensure complete and homogeneous paste dispersion. Then, aliquots of 100 μL are taken for each substrate and deposited onto the c-TiO$_2$ layer by spin-coating (Laurell, model WS-400BX-6NPP Lite). This step involves a uniform spread of the 100 μL solution over the substrates and spinning at 4000 rpm (acceleration of 2000 rpm/s) for 10 seconds. After spin-coating, samples are immediately transferred to a heating plate (Gestigkeit, hotplate model PZ28-3TD, PID-Temperature Controller model PR3-3T) and heated to 100 °C for 10 minutes. Then, the samples are further annealed with a series of ramps and holding temperatures: i) ramp at 25 ºC·min$^{-1}$ to 125 ºC during 5 minutes, ii) ramp at 13 ºC·min$^{-1}$ to 325 ºC during 5 minutes, iii) ramp at 10 ºC·min$^{-1}$ to 375 ºC during 5 minutes and iv) ramp at 15 ºC·min$^{-1}$ to 450 ºC during 30 minutes. Once cooled, the resulting samples are treated with UV/O$_3$ for 15 minutes.

*2.5 Deposition of Organometal Halide Perovskite film.*

For the synthesis process described below, the following high-purity reagents and solvents were used: Lead iodide (PbI$_2$), TCI, purity 99.99%; Formamidinium iodide (FAI, low water content), TCI, purity > 99.0%; Lead bromide (PbBr$_2$), TCI, purity > 98.0%; Methylammonium bromide (MABr, low water content), TCI, purity > 98.0%; Dimethyl sulfoxide (DMSO), Acros Organics, > 99.9%; N,N-Dimethylformamide (DMF), Acros Organics, purity> 99.8%; Cesium iodide (CsI), Alfa Aesar, purity > 99.998%; Rubidium iodide (RbI), Merck, purity 99.9%;



Once the electron transport layers have been deposited, samples are transferred to a nitrogen-filled glovebox, where the temperature is maintained between 24 and 27 ºC since temperatures above 30 ºC are known to cause perovskite degradation.[10] A precursor solution is prepared by mixing two mother solutions to prepare the perovskite film. The first mother solution contains $PbI_2$ and FAI at 1 M concentration dissolved in DMSO:DMF in a 1:4 vol% ratio. The second solution includes $PbBr_2$ and MABr in the same concentration and solvent. These two mother solutions are mixed in a 5:1 vol% ratio, respectively, to prepare the first mixture, known as MAFA solution. Two additional precursor solutions are prepared, CsI and RbI, both at 1.5 M, in DMSO and DMSO:DMF in a 1:4 vol% ratio, respectively. First, a 5 vol% of CsI is added to the MAFA solution (named CsMAFA). Then, the quadruple cationic composition (RbCsMAFA) is prepared by adding 5 vol% of the RbI solution to the previous CsMAFA mixture. For the perovskite film, 120 μL of RbCsMAFA solution is spread evenly over the samples for spin-coating, consisting of two stages: 1) at 1000 rpm for 10 s and 2) at 6000 rpm for 20 s. During the latter step, 200 μL of chlorobenzene is added to the centre of the sample, acting as an anti-solvent 5 seconds before the end of the program. Immediately after the spin-coating, the samples are placed on a heating plate at 100 ºC for 60 minutes.

*2.6 Deposition of Hole Transporting Layer (HTL).* For the synthesis process described below, the following high-purity reagents and solvents were used: (N2,N2,N2′,N2′,N7,N7,N7′,N7′-octakis(4-methoxyphenyl)-9,9′-spirobi[9H-fluorene]-2,2′,7,7′-tetramine) (Spiro-MeOTAD), Merck, purity 99%; Chlorobenzene, Acros Organics, purity 99.5%; Bis(trifluoromethane)sulfonimide lithium salt (LiTFSI), Merck, purity 99.95%; Acetonitrile, Acros Organics, purity 99.9%; tris(2-(1H-pyrazol-1-yl)-4-tert-butylpyridine)cobalt(III)tris(bis(trifluoromethylsulfonyl)imide) (FK209 Co(III) TFSI salt), Merck, purity 98%; and 4-tert-Butylpyridine, Merck, purity 98%.

A 0.07 M solution of Spiro-MeOTAD in chlorobenzene is prepared for the hole-selective contact layer. Separate solutions of LiTFSI (1.8 M in acetonitrile) and FK209 (0.25 M in acetonitrile), are also prepared. To dope the Spiro-MeOTAD solution, specific volumes of the LiTFSI and FK209 solutions, along with 4-tert-butylpyridine, are added to achieve final molar concentrations of 0.5 M, 0.03 M, and 3.3 M for LiTFSI, FK209, and 4-tert-butylpyridine, respectively. Then, 50 μL of the resulting solution is spread onto the perovskite, followed by a spin-coating process at 4000 rpm for 15 s. This achieves a homogeneous distribution of the spiro-MeOTAD film over the substrate.

*2.7 Evaporation of gold metal electrode.* Before gold deposition, the selective contacts (m-$TiO_2$ and HTL) and perovskite layer are removed from part of the substrate to facilitate direct



placement of the gold contact on FTO, ensuring good electrical contact. Subsequently, a mask is placed to define the electrodes (each electrode has an area of 1.5 x 0.3 cm$^2$) and transferred to the gold evaporator. Finally, an 80 nm gold layer is deposited by thermal evaporation in a high vacuum system at a pressure below 10$^{-5}$ mbar.

*Characterization Methods.* High-resolution scanning electron microscopy (SEM) images of the samples were acquired using a Hitachi S4800 microscope operating at 2 kV. Cross-sectional views were obtained by cleaving the substrates. Transmission electron microscopy (TEM) images were captured using a CM20 apparatus from Philips. High-angle annular dark-field scanning transmission electron microscopy (HAADF-STEM) images were acquired using a Tecnai G2 F30 S-Twin STEM and Titan Themis from FEI. Rutherford backscattering spectroscopy (RBS) and nuclear reaction analysis (NRA) characterizations were conducted at the 3 MV Tandem Accelerator of the National Centre of Accelerators (Seville, Spain). RBS measurements were performed with α-particles of 2.0 MeV and a passivated implanted planar silicon (PIPS) detector located at a 165° scattering angle. NRA was utilized to determine the presence of C, N, and O elements in the film through the $^{12}$C(d,p)$^{13}$C, $^{14}$N(d,α$_1$)$^{12}$C, and $^{16}$O(d,p$_1$)$^{17}$O nuclear reactions using deuterons of 1.0, 1.4, and 0.9 MeV, respectively. The spectra were obtained using a particle detector at a 150º collection angle in combination with a 13 μm thick Mylar filter to intercept the backscattered particles. NRA and RBS spectra were simulated using the SIMNRA 6.0 code. Film densities were determined from the combined RBS and NRA analyses, with thickness values obtained from cross-sectional SEM micrographs of the films. XPS characterizations were performed in a Phoibos 100 DLD X-ray spectrometer from SPECS. Before analysis, samples underwent pre-treatment via Ar plasma ablation (10$^{-6}$ mbar, 5 kV, 5 minutes, 3 mA). The spectra were collected in constant pass energy mode, with 50 eV for the survey spectrum and 35 eV for the high-resolution zone spectra, using an Mg Kα source. C1s signal at 284.8 eV was utilized to calibrate the binding energy (BE) in the spectra due to surface charging during measurements. The assignment of the BE to the different elements in the spectra corresponds to the data in the references.[22–25] Optical transmittance spectra in the wavelength range of 200-2500 nm were recorded in a PerkinElmer Lambda 750 S UV–Vis–NIR spectrophotometer on samples deposited of fused silica substrates. An effective medium model (EMA) approximation, concretely Bruggeman model, was utilized to determine the effective refractive index of each sample. This model establishes a relationship between the refractive index (n) of the porous material and its porosity, assuming a homogeneous distribution of pores within the material.[26] Considering the possibility of pores being filled with either air ($n_a$ = 1.00) or water ($n_w$ = 1.33), the Bruggeman Model allowed for a



comprehensive assessment of the refractive index variation, providing valuable insights into the optical properties of the samples. Specifically, for $TiO_2$, the refractive index of the bulk material depends on the crystalline structure. As it will be discussed, the porous structure synthesized is essentially amorphous, with inclusions of small crystals of anatase or rutile. Thus, we have chosen an intermediate refractive index of anatase ($n_{amorph}<n_{anatase}<n_{rutile}$) of $n_{ox}$=2.65 (@550 nm)[27]. The Bruggeman equation can be expressed as follows[28]:

$$ff_{ox}\frac{n_{ox}^2-n_{eff}^2}{n_{ox}^2+2n_{eff}^2} + ff_a\frac{n_a^2-n_{eff}^2}{n_a^2+2n_{eff}^2} + ff_w\frac{n_w^2-n_{eff}^2}{n_w^2+2n_{eff}^2} = 0 \qquad \text{(Equation 1)}$$

where $n_{eff}$ represents the effective refractive index of the layer, and $ff_{ox}$, $ff_a$ and $ff_w$ are the volume fractions of the oxide, air, and water, respectively. This equation provides a mean to calculate the effective refractive index of the composite material, incorporating the contributions from each constituent phase and their respective volume fractions.

Current density-voltage (J-V) curves were recorded under a solar simulator (ABET-Sunlite) with an AM 1.5G filter at 100 mW·cm$^{-2}$. These curves were measured using a black mask of 0.14 cm$^2$ to define the active area. Electrochemical impedance spectroscopy (EIS) was carried out using a potentiostat module (PGSTAT302N/FRA2, Autolab) following our previously described methodology[29]. Specifically, the EIS experiments were conducted by applying a 20 mV perturbation in the frequency range of $10^6$ - 0.1 Hz at open-circuit conditions. The open-circuit photovoltage was obtained by varying the light intensity using red ($\lambda$= 635 nm) and blue ($\lambda$= 465 nm) light-emitting diodes (LEDs) over a wide range of DC light intensities.

## 3. Results and discussion

The plasma-assisted deposition of metal-containing small-molecules and subsequent plasma etching methodology consists mainly of two steps:[30–32] first, the metalorganic molecules are deposited by remote plasma-assisted vacuum deposition (RPAVD step), and second, the samples are subjected to a soft plasma etching (SPE) at low-mild temperatures (see **Scheme 1**) (see Experimental section for details). It is important to note that, during the RPAVD step, plasma composition is a determining factor in the microstructure of the layers, as shown in the SEM micrographs in **Figure 1**. While the deposition under argon plasma (RPAVD-Ar*) produces continuous and flat films (panel a)), the use of oxygen plasma (RPAVD-$O_2$*) induces nanocolumnar growth (panel b)). Hence, the electrodes prepared by RPAVD-$O_2$* and RPAVD-$O_2$*+SPE are called *Columnar* and *Columnar+SPE*, respectively. *Columnar-Aerogel $C_1$* and *Columnar-Aerogel $C_2$* samples refer to those electrodes prepared by RPAVD-Ar*+SPE (one or two cycles, respectively) on top of the *Columnar+SPE* layers. We have previously reported that repeating these two processes (RPAVD-Ar*+SPE) sequentially makes fabricating aerogel-like ultraporous films with controlled thicknesses possible.[32] It can be noted that the *Columnar* sample presents a thin and compact layer of ca. 20-30 nm at the interface with the substrate. This interfacial compact layer has been extensively reported for columnar growth in vapor deposition methods. In the case of *n-i-p* or *p-i-n* architectures for optoelectronic devices, such as the solar cells presented here, this naturally growing compact interfacial layer is very convenient to selectively promote electrons or holes injection at the electrode, reducing the recombination at the contact and thus increasing the efficiency of the devices.

The subsequent SPE treatment of both samples induces a complete modification in the microstructure of the layers, as shown in Figure 1 c) for RPAVD-Ar* and d) for RPAVD-$O_2$*. In both cases, the thickness of the films is significantly reduced to around 40% of the original.[32] The microstructural modification can be easily observed in the surface views of Figures 2 c-d). While the SPE treatment on the RPAVD-Ar* samples produces a highly porous interconnected network with a broad distribution of porous structures in the range from 10 to 250 nm and maximum in the region near 20 nm (as reported previously in [32]), the *Columnar* sample maintains the initial columnar microstructure with a higher degree of coalescence, but with much smaller open pore structures than the previous case, mean size below 80 nm, with maxima in the region of 10 nm (See Figure S1).

In this work, we have developed an ETL using a combination of RPAVD-$O_2$* and -Ar* layers, consisting of I) RPAVD-$O_2$*+SPE (*Columnar+SPE*), II) RPAVD-Ar*+SPE (*Columnar-Aerogel $C_1$*), and III) a second cycle of RPAVD-Ar*+SPE (*Columnar-Aerogel $C_2$*) as it is shown in Scheme 1. The final ETL architecture of *Columnar-Aerogel $C_2$* samples is shown in Figures 2 e) and f). It can be noted how the surface view of this sample consists of a highly porous and homogeneous interconnected network. The higher magnification image (f) shows the appearance of bigger porous structures and voids compared to the direct deposition on flat substrates (Figure 1 c). These big voids are surrounded by a network containing very small pores, as observed in the inset of Figure 1 f). From a cross-section image of *Columnar-Aerogel $C_2$* sample (Figure 1 f), it is observed how two cycles of RPAVD-Ar*+SPE produce the enclosing of the columns but provide a porous structure that will later enable the infiltration of the perovskite.

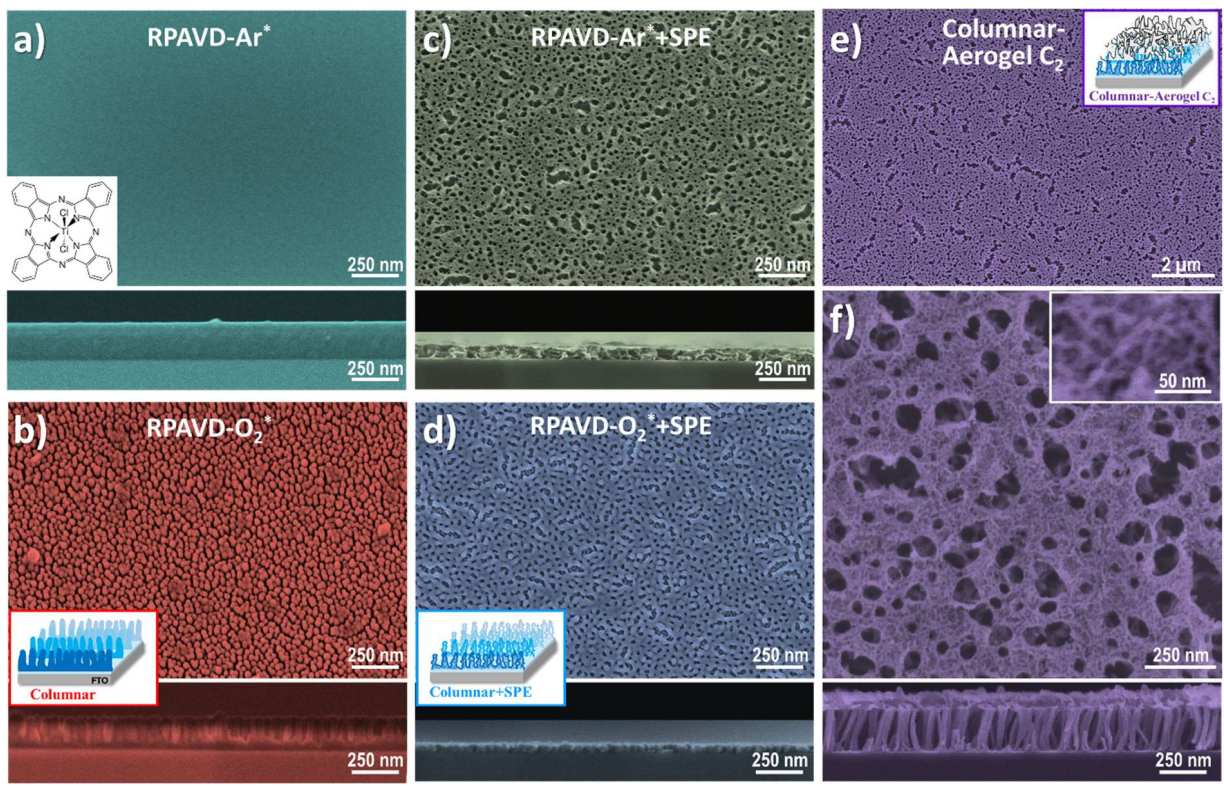

**Figure 1.** Surface and cross-sectional SEM Images of RPAVD-Ar* (a,c) and -O$_2$* (b,d) before (a,b) and after the oxygen SPE (c,d) treatment. e-f) Surface and cross-sectional SEM Images of the final ETL architecture (*Columnar-Aerogel C$_2$*).

*Columnar-Aerogel C$_2$* sample was further examined by STEM (**Figure 2**), confirming the structure observed by SEM. That is, columns from RPAVD-O$_2$* merged at the tips by a porous network from RPAVD-Ar*. These samples comprise cavities with walls about 3 nm thick, as observed in the high-resolution (HRTEM) image shown in Figure 2 b). Besides, It can be seen in the micrograph that the columns are also porous. Moreover, the HRTEM image and its corresponding digital diffraction pattern (DDP) (Figure 2 b) show the formation of small nano-crystalline domains embedded in an amorphous structure. The interplanar spacing values measured of 2.4 Å and 2.5 Å, could be assigned to the (004) family plane of the anatase phase and (101) of the rutile phase, while the value of 2.1 Å fits well with the (111) family planes of the rutile phase. The development of a crystalline phase, especially the rutile in TiO$_2$, is a very striking result since the temperatures involved in the synthesis are very far away from the required for the crystallization of this oxide, which usually requires temperatures above 600 ºC in the case of the rutile phase.

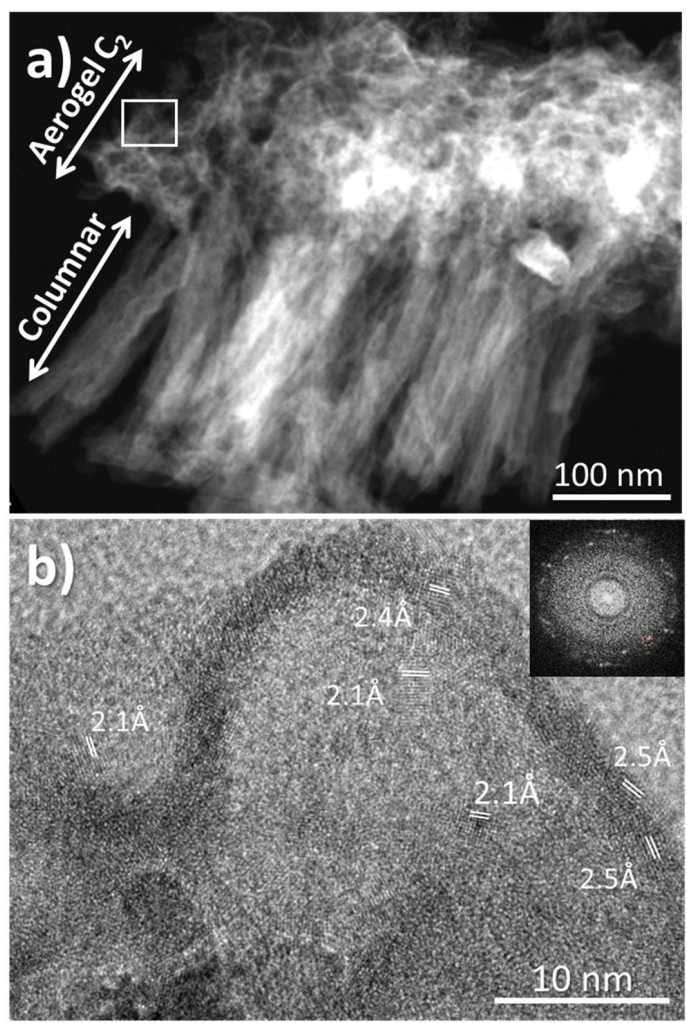

**Figure 2.** a) High Angle Annular Dark Field Scanning -TEM (HAADF-STEM) image showing the nano-structure of the *Columnar-Aerogel $C_2$* sample. b) High-resolution TEM image of the square area marked in a) and its corresponding digital diffraction pattern (DDP) as inset. Figure b) also shows the main interplanar distances identified in the image.

Defining the chemical composition and stoichiometry of these ETLs is not straightforward, as they have been obtained after plasma-assisted decomposition of the metalorganic precursors. Together with their high porosity, they can promote rich hydroxylate or carbonate surfaces that might affect the crystallinity, efficiency in electron transport and collection, and stability of the perovskite/$TiO_2$ interface. Thus, in **Figure 3**, we gather a thorough XPS analysis of the different synthesis steps, which also incorporates a spin-coated mesoporous (m-$TiO_2$) sample annealed at 450ºC as reference (see experimental section for further details). The elements present in the spectra are Ti, O, C, and N, the latter coming from the TiPcCl molecule. The survey spectra and the atomic/relative percentages (left/right axes) of every element are shown in Figure 3 a) and b), respectively. It can be noted that none of the samples presents any trace of chlorine coming from the molecule. This dichlorination has been reported after oxygen plasma treatment of other metalorganic molecules due to the metal oxidation (in this case, Ti) and formation of volatile chlorine species such as ClH and $Cl_2$.[33] Besides, the *Columnar* layer (before SPE treatment) differs in the atomic percentages (Figure 3 b) as expected from the stoichiometry of the TiPcCl molecule with formula $C_{32}H_{16}Cl_2N_8Ti$. The relative amounts of N and C from the XPS analysis with respect to Ti are $(N:Ti)_{XPS}=0.74$ and $(C:Ti)_{XPS}=2.54$, which is more than one order of magnitude below the expected from the formula, which are $(N:Ti)_{formula}=8$ and $(C:Ti)_{formula}=32$, indicating that the RPAVD in an oxygen plasma produces the oxidation of the TiPcCl molecule, forming volatile carbon and nitrogen byproducts. The SPE treatment of this *Columnar* layer further reduces the amount of nitrogen (below 1% in atomic percentage) and the carbon to 18% ($(C:Ti)_{XPS}=0.82$). In addition, the SPE-treated sample presents an excess of oxygen ($(O:Ti)_{XPS}=2.72$). The excess carbon and oxygen can be attributed to surface contamination, such as adsorbed hydrocarbons and water molecules in the highly porous oxide surfaces. Note that the reference m-$TiO_2$ sample also presents an excess of carbon and oxygen ($(C:Ti)_{XPS}=0.55$ and $(O:Ti)_{XPS}=2.40$), but those values are slightly lower, as can be expected due to the lower porosity of the reference sample. However, a small percentage of carbon arising from an incomplete oxidation of the RPAVD layer cannot be discarded. *Columnar-Aerogel $C_1$* and $C_2$ samples do not significantly alter the relative amount of any element.

Figure 3 c) shows the XPS spectra of the Ti2p region of the samples synthesized in this work, compared to the reference m-$TiO_2$ samples. For all samples, two peaks at 458.5 and 464.2 eV corresponding to $Ti2p_{3/2}$ and $Ti2p_{1/2}$, respectively, have been deconvoluted. These results confirm that $Ti^{4+}$ is the only species of titanium present, regardless of experimental variations. This observation is characteristic of $TiO_2$ and aligns with the well-known binding energy (BE) values for $Ti^{4+}$ in this phase.[22] It is worth noting that the samples have been *in*-situ bombarded with Ar ions to verify that $Ti^{4+}$ is also present along the depth and not only at the surface. The C1s spectra (Figure 3d) show two main peaks: saturated C-H at 285.0 eV and unsaturated C=C at 284.5 eV. Additional peaks appear at 286.5 eV (C-O, linked to hydroxyl groups), 288.0 eV (C=O and imine-type C=N), and 289.0 eV (O-C=O, indicating carboxylic groups). Before SPE treatment, *Columnar* samples exhibit strong 288.0 eV and 289.0 eV signals, suggesting incomplete oxidation of the TiPcCl precursor and preferential carboxylate formation. The deconvoluted N1s region supports this phenomenon (see Supporting Information S2). After SPE treatment (*Columnar*+SPE, *Columnar-Aerogel $C_1$*, and $C_2$), samples show dominant C=C and C-H signals, with minor C-O, C=O, and O-C=O peaks. Their carbon profile is closer to

m-TiO$_2$ but with a higher C=C/C-H ratio, indicating slight chemical variations that may influence their properties and applications.

The zone spectra of O1s are shown in Figure 3 e). For all samples, three peaks have been identified at binding energies of 529.6 eV, 530.7 eV, and 531.8 eV, usually labeled in literature as O$_M$, O$_V$, and O$_H$.[6,34–37] The peak at 529.6 eV (O$_M$) is attributed to O$^{2-}$ ions combining with the Ti atom in a stoichiometric TiO$_2$ structure. The peak at 530.7 eV (O$_V$) is related to surface oxygen deficiency, which may result from incomplete film oxidation during the deposition process or from impurities. Since the Ti2p peaks show no signs of Ti$^{3+}$ present, we can only attribute this peak to surface contamination. The peak at 531.8 eV (O$_H$) is assigned to the hydroxyl groups anchored to the surface. High oxygen vacancy (O$_V$) and hydroxyl (O$_H$) peak intensities in the O1s deconvoluted spectra are commonly associated with a higher fraction of amorphous TiO$_2$. This is because amorphous TiO$_2$ exhibits higher oxygen vacancies and hydroxyl group density than its crystalline counterpart. These defect-related states, typically observed at binding energies around 530.5–532.0 eV, indicate a more disordered structure, as already observed in low-temperature processed TiO$_2$ films.[6,34–37] The O$_H$ region also encompasses lower-intensity peaks attributable to the presence of other oxygen-containing species, such as C=O and C-O residues, as mentioned above. However, its deconvolution was unsolvable due to its low concentration.

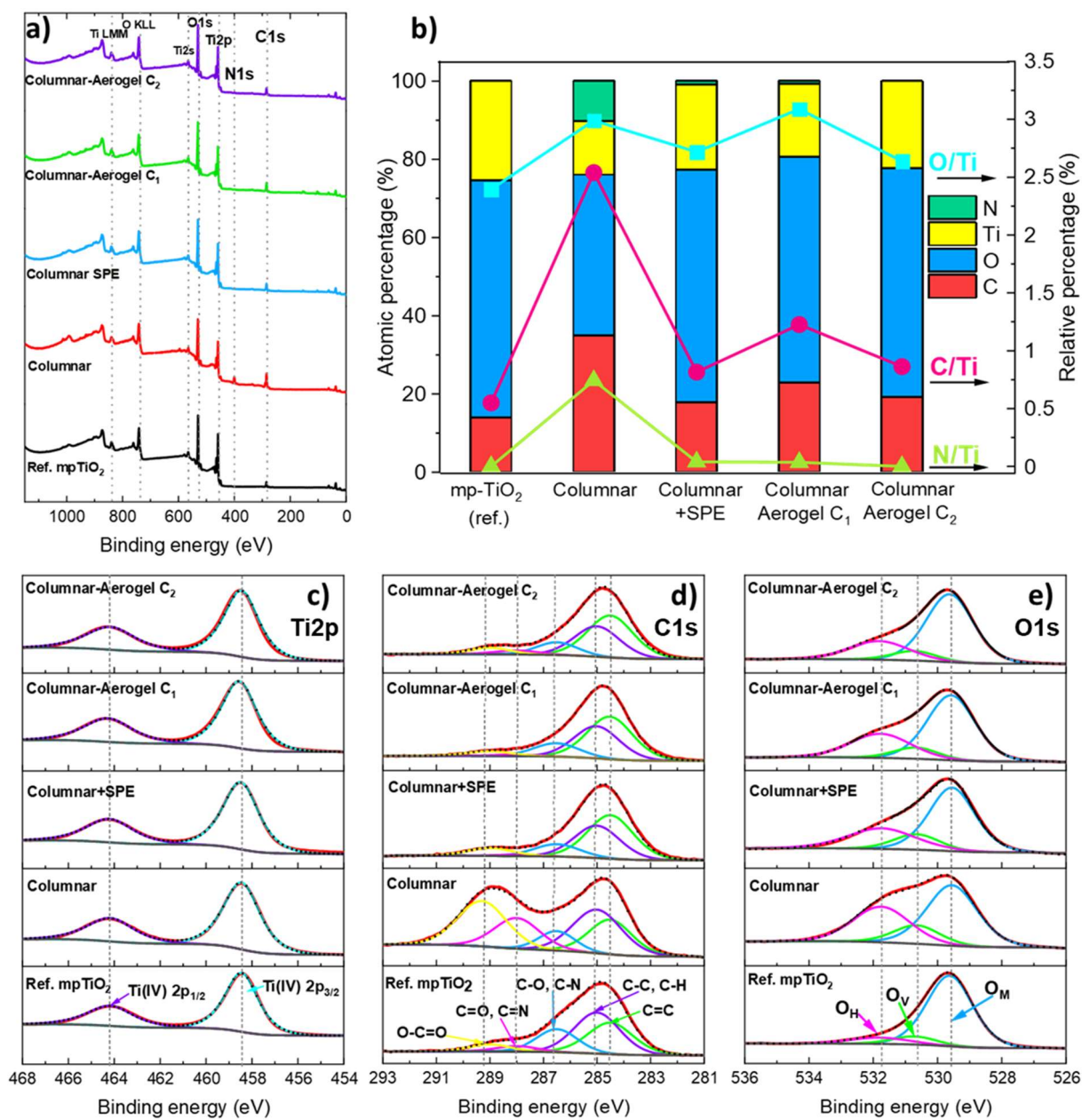

**Figure 3**. XPS survey (a) and detailed zone spectra of Ti2p (c), C1s (d), and O1s (e) of the RPAVD-$O_2$* before (*Columnar*) and after SPE treatment (*Columnar* + SPE) samples, as well as *Columnar Aerogel $C_1$* and *$C_2$*. b) Atomic (left axis) and relative atomic percentages (right axis) quantified from the XPS spectra.

**Figure 4** presents the area of the $O_M$, $O_V$, and $O_H$ peaks normalized to total oxygen ($O_T$= $O_M$+$O_V$+$O_H$) for each sample type. In the *Columnar* samples, the $O_M$, $O_V$, and $O_H$ areas were 51.8%, 14.5%, and 33.7%, respectively. Although the fragmentation of the TiPcCl molecules occurs in an oxygen plasma environment, the full oxidation to $TiO_2$ is not complete, showing the highest concentration of surface oxygen defects ($O_V$=14.5%), that together with a high $O_H$ ratio suggests the presence of amorphous regions in the films.[6,34–37] When these samples are subjected to the SPE process (*Columnar+SPE*), the $O_M$ area increases to 58.1%, while the $O_V$ and $O_H$ areas decrease to 13.9% and 27.9%, respectively. As expected, the plasma etching process reduces defects related to oxygen vacancies and improves the crystallinity of the $TiO_2$ film. The effect is much more pronounced for *Columnar-Aerogel* samples, with an increased $O_M$ area of 61.6% and 68.0% for $C_1$ and $C_2$, respectively, while the $O_V$ area is further decreased to 9.8% and 8.8%. This latter value is even lower than the ratio measured for reference samples, with an $O_V$ area value of 9.1%. These results indicate that the reported methodology is very efficient in synthesizing low-defect crystalline $TiO_2$ layers with similar characteristics to those formed at high temperatures. Moreover, the addition of the final layers of the RPAVD-Ar*+SPE layers (*Columnar-Aerogel*) produces a decrease in the $O_H$ area to 28.6% and 23.1% for $C_1$ and $C_2$, respectively, indicating a lower presence of hydroxyl surface species. However, the high-temperature processed $TiO_2$, shows a significantly lower $O_H$ area of 9.3%, suggesting a slightly lower crystallinity of the films synthesized with our plasma methodology.[6,34–37]

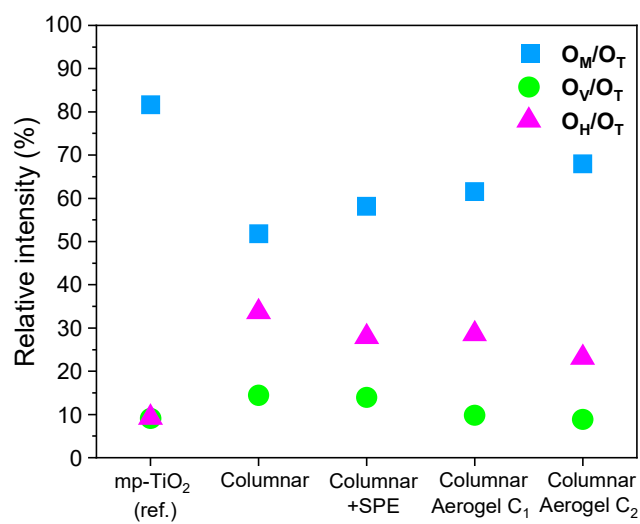

**Figure 4.** Areas of the different peaks deconvoluted from the detailed XPS zone of O1s: $O_M$ (blue), $O_V$ (green), and $O_H$ (magenta) normalized to total oxygen ($O_T = O_M+O_V+O_H$) for each sample type.

**Figure 5** shows the transmittance spectra in the UV-Vis-NIR region of all the studied layers, as well as the m-$TiO_2$ reference and a fused silica substrate. The fused silica substrate presents a highly flat spectrum at 93% transmittance above 400 nm. The reference m-$TiO_2$ sample shows an oscillatory profile in the Vis-NIR region, which maximum is modulated by the fused silica transmittance level. As usually occurs for high refractive index nanoparticulate coatings, the layers also present scattering, noticeable from lowering the transmittance values as the wavelength decreases. By contrast, the oscillations are highly attenuated for the *Columnar* (red) and *Columnar+SPE* (blue) samples. The reduction in the number of oscillations, along with the non-zero transmittance at wavelengths below the titanium dioxide bandgap (ca. 350 nm), which is commonly observed in compact titanium dioxide layers thinner than 50 nm, suggests that the films have high porosity. This increased porosity brings the refractive index of the layers closer to that of fused silica. The density and porosity characteristics of the layers are depicted in the inset of Figure 5. Specifically, for the *Columnar* samples before and after the SPE treatment, the porosity values are recorded as 78.5% and 79.8%, respectively. The *Columnar-Aerogel $C_1$* and $C_2$ show significantly increased transmittance, with values well above the fused silica substrate. This behavior occurs for antireflective coatings, which are layers with a smaller refractive index than the substrate, in this case, fused silica with n=1.459 @550 nm.

Figure 5 b) shows the theoretical refractive index of $TiO_2$ films as a function of the porosity considering an Effective Medium Approximation (see Experimental Section for further details). Please note that the graph displays two curves (solid lines), one considering that the pores are empty (black curve) and the other with the pores filled by water (blue curve). The vertical dotted lines indicate the porosity value displayed in the inset of Figure 5 a) for the different samples, i.e. *Columnar* (red), *Columnar+SPE* (blue), *Columnar-Aerogel C1* (green), and *C2* (purple). At the same time, the X highlights the refractive index, considering that 80% of the pores are filled with water. It can be noted that the *Columnar* samples before and after SPE would present an antireflective behavior if the pores are empty (see the intersection between the black curve and the vertical red and blue dotted lines at the corresponding porosity values). However, the transmittance values lower than the substrate ones over the entire Vis-NIR range would indicate that the pores are partially filled, presumably with condensed water from the atmospheric moisture, thus increasing their effective refractive index. The subsequent layers (*Columnar-Aerogel $C_1$ and $C_2$*) drastically increase porosity, reaching 85.8% (see the inset of Figure 5 a)). As it can be observed in Figure 5 b), even assuming that a portion of the pores are filled with water (the figure shows an example with 80% of the pores filled with water), it is reasonable that the effective refractive index is below the fused silica substrate (yellow horizontal line) for the *Columnar-Aerogel* cases. According to the literature, a highly porous ETL can improve photovoltaic performance by enhancing perovskite infiltration and increasing surface area. This facilitates efficient electron transport, reduces charge accumulation, and improves charge extraction while minimizing recombination losses.[38–40] Therefore, the plasma-based ETLs reported here are expected to impact the efficiency of perovskite solar cells positively.

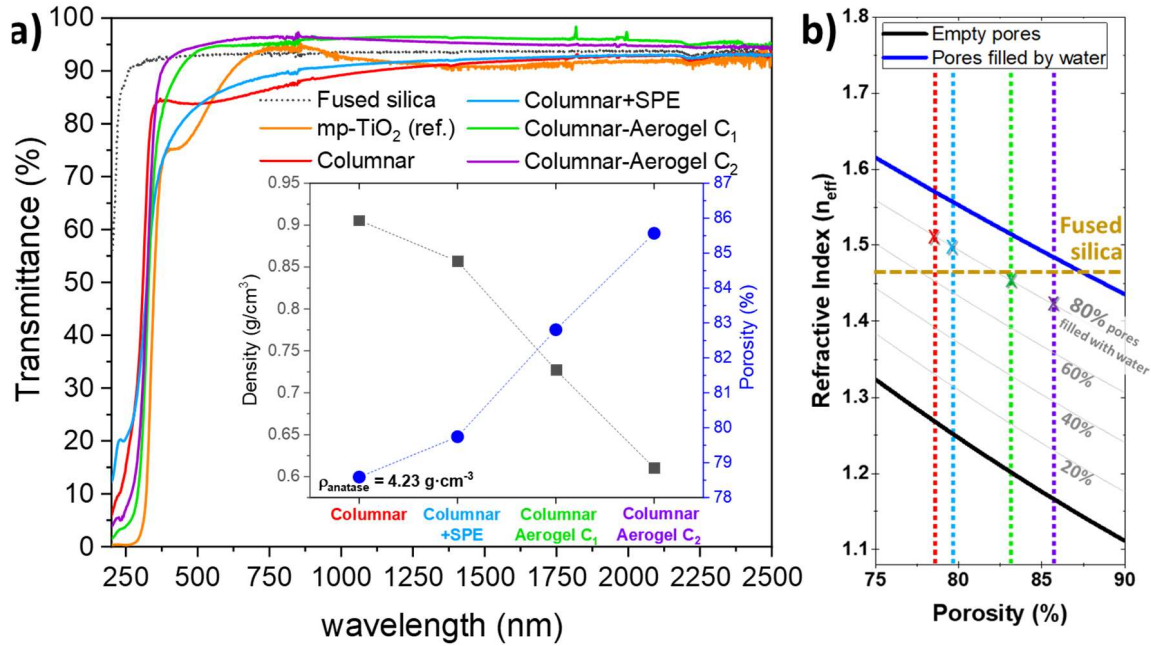

**Figure 5.** a) Transmittance spectra of all the samples studied, including an m-TiO$_2$ reference sample and the bare fused silica substrate. The inset shows the samples' density and porosity (left and right axes, respectively). b) Effective refractive index of TiO$_2$ films as a function of the porosity considering the Bruggeman model as an Effective Medium Approximation. The black and blue curves indicate the effective refractive indexes, considering that the pores are empty and filled with water. The vertical dotted lines indicate the porosity value displayed in the inset of the panel a) for the *Columnar* (red), *Columnar+SPE* (blue), *Columnar-Aerogel C$_1$* (green), and *C$_2$* (purple) samples. The X displays the refractive index, considering that 80% of the pores are filled with water

The TiO$_2$ plasma-based coatings have been employed as electron transport layers in a direct *n-i-p* perovskite solar cell. **Figure 6** a) shows the champion current density-voltage curves (J-V curves) measured under 1-sun AM1.5G illumination for the different samples, and b-e) present the statistical study of the photovoltaic (PV) parameters (V$_{OC}$ (b), J$_{SC}$ (c), FF(d), and PCE (e)). A standard, fully optimized perovskite solar cell containing compact- and mesoporous TiO$_2$ layers annealed at 450 ºC has been added for comparison (reference). From the series of studied cells, the lowest performance was obtained for the *Columnar* sample (before the SPE treatment), mainly due to a poor short-circuit current density (J$_{SC}$) (see Figure 6 c). This sample also showed the highest dispersion in all the photovoltaic data. By contrast, the SPE treatment (*Columnar+SPE*) (blue) significantly improves photovoltaic behavior as J$_{SC}$ increases considerably. An improvement in J$_{SC}$ was also observed for the subsequent layers *Columnar-Aerogel* as the number of cycles increased, which is also consistent with a higher PVK thickness for the *Columnar-Aerogel C$_2$* samples (see Figures S3 and S4). The increased J$_{SC}$ resulted in a substantial improvement in PCE and a decreased dispersion in the PV parameters. The most significant efficiency boost was achieved with *Columnar-Aerogel C$_2$*, reaching 13.0 ± 1.6% (compared to the reference average of 16.6 ± 0.9%) while maintaining a comparable open-circuit voltage (V$_{OC}$) to the reference samples. The J$_{SC}$ and fill factor (FF) were slightly lower than the reference samples. Table S1 presents the photovoltaic parameters obtained from these J-V curves for each configuration studied.

The architecture of the low-temperature plasma layers, consisting of a columnar TiO$_2$ structure and an aerogel-like layer on top, has been rationally designed to maximize the photovoltaic response. In this work, the most efficient devices implementing low-temperature plasma layers (*Columnar-Aerogel C$_2$*) consist of a first columnar structure of 250 nm followed by 100 nm of aerogel-like TiO$_2$, with a total thickness of ca. 350 nm. This thickness is higher than the standard reference solution-based ETL, consisting of a c- and m-TiO$_2$ reference device with 250 nm. Cross-section images of both devices are shown in Figure S3, where it can be noted a slightly different thickness perovskite of 700 and 800 nm for the reference and *Columnar-Aerogel C$_2$*, respectively, which, in both cases, are higher than the ETL (see also Figure S4 for the ETL+PVK thickness). The choice of the first columnar layer for the perovskite solar cells is double. In the first place, the continuous TiO$_2$ thin layer growing at the FTO interface (see cross-section image in Figure 1 f) plays a similar role to the high-temperature c-TiO$_2$ layer in reference devices, avoiding the direct contact of the perovskite layer with the FTO reducing hole recombination at this electrode. In the second place, the open space between the columns allows the growth of the aerogel-like layers on top with more open and broader voids than one grown on flat substrates (the comparison of both surface views is shown in Figure S5). Thus, the perovskite can easily diffuse inside the columnar voids, maintaining proper connectivity on the perovskite from the thin TiO$_2$ layer at the FTO interface to the Spiro-OMeTAD. The presence of only one of the low-temperature plasma-synthesized TiO$_2$ layers on top of FTO, either columnar or aerogel-like, depicts much lower photovoltaic behavior. This can be seen in Figure 6 b-e) for the *Columnar+SPE* or the two additional sets of values in the rectangle in panel e) corresponding to *Aerogel C$_1$* and *C$_2$* (without the *Columnar* layer). These devices show significantly lower photovoltaic parameters than the full one containing columnar and aerogel-like layers (see also Table S1).

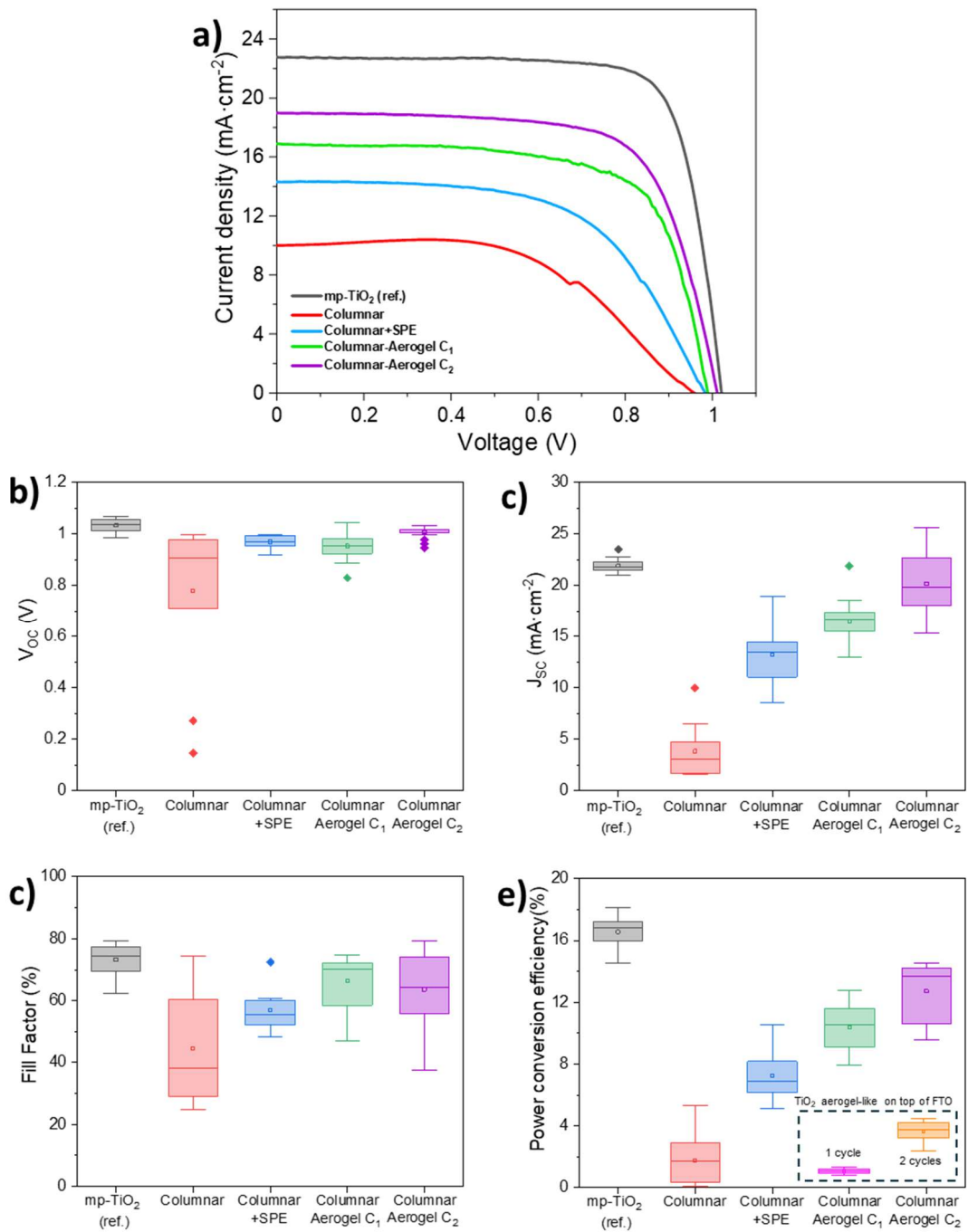

**Figure 6.** a) Current density-voltage curves of champion devices measured under 1-sun AM1.5G illumination incorporating the different plasma-based ETLs and a high-temperature $TiO_2$ (compact + mesoporous) reference. b-e) Statistical analysis of the different photovoltaic parameters: $V_{OC}$ (b), $J_{SC}$ (c), FF (d), and PCE (e). This latter figure also incorporates *Aerogel $C_1$* and *$C_2$* devices directly deposited on FTO (without *Columnar* layer).

Impedance spectroscopy (IS) at open-circuit conditions was performed to analyze the impact of plasma-based ETLs on the electronic dynamics of PSCs. The voltage has been fixed by LED illumination using light sources with two different wavelengths and thus different light penetration into the perovskite absorbing layer: 635 nm (red light) and 465 nm (blue light). Due to the high absorption of perovskite in the blue region, the 465 nm LED light is absorbed near the TiO$_2$/perovskite interface. By contrast, the 635 nm illumination, due to its higher penetration depth, allows the characterization of the electron charge behavior in the perovskite region close to the HTL.[41] The IS spectrum for a PSC is typically characterized by two signals (both are coupled): the high-frequency (HF, $10^6$-$10^3$ Hz) associated with the electronic transport and recombination processes, and the low-frequency signal (LF, < 10Hz) affected by ionic accumulation and migration.[41–44] Focusing on the HF signal, **Figure 7** shows the time constant (a) and recombination resistance (b) for the blue and red illumination wavelengths (with blue and red colors, respectively) as a function of the open-circuit photovoltage extracted by fitting the IS response to a simple Voight equivalent circuit (inset in Figure 7 b). No significant differences are obtained between blue and red excitations for any of the devices analyzed, confirming that our plasma-based ETLs do not introduce any preferential recombination site.[41] This aligns with the XPS data (Figure 3) and conclusions from Figure 4, which indicate that the plasma-synthesized TiO$_2$ films exhibit high crystallinity (also confirmed by TEM in Figure 2) with low oxygen vacancy (O$_V$) concentrations and no detectable Ti$^{3+}$ species. Although their crystallinity is slightly lower than the high-temperature reference, the plasma method yields highly crystalline, low-defect TiO$_2$ layers with comparable structural quality, explaining the absence of additional recombination sites.

This result is also consistent with the almost constant value of V$_{oc}$ obtained for *Columnar-Aerogel C$_1$* and *C$_2$*, and reference devices (see Figure 6 b). However, comparing the R$_{HF}$ and C$_{HF}$ (charge recombination resistance and capacitance at HF, respectively) under red light with respect to the photovoltage (Figure S6), a different behavior is observed for the PSC with plasma-based ETLs. In particular, a lower slope is obtained in the R$_{HF}$ for plasma-based devices (Figure S6 a), suggesting that the main mechanism of recombination resistance has been altered by introducing the *Columnar Aerogel* electrodes with respect to the trap-limited recombination mechanism observed for the reference sample. Since the R$_{HF}$ obtained for red and blue signals are almost coincident (Figure 7 b), the variations in the slopes can be explained by the different electrical features that perovskite can show depending on the nature of the ETL, i.e. crystallization, grain size, etc. On the other hand, the C$_{HF}$ values are almost constant for all the devices over the bias range analyzed (Figure S6 b), which agrees with the geometrical capacitance of perovskite.[44]

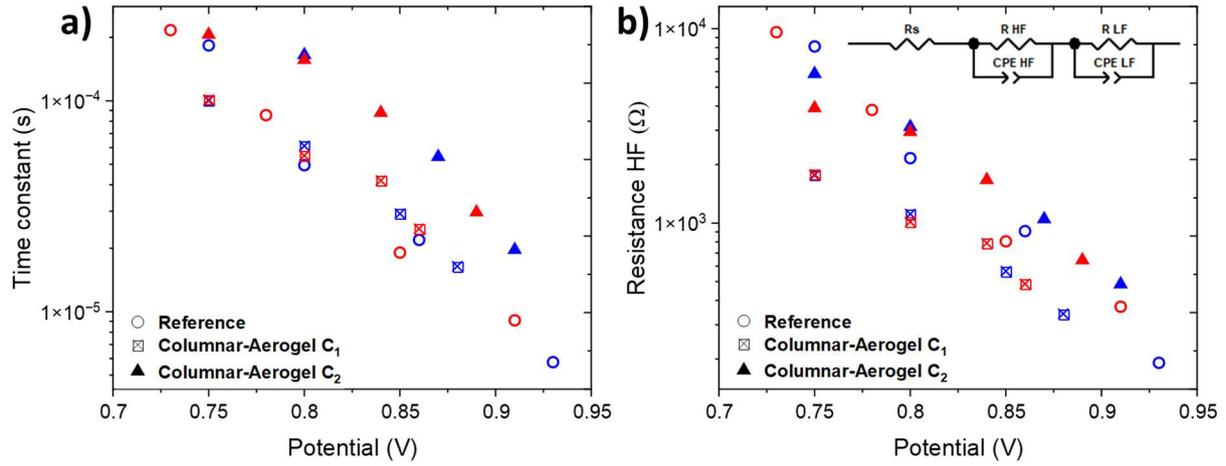

**Figure 7.** High-frequency time constant (a) and resistive elements (b) extracted from fittings of the impedance spectra obtained at open-circuit conditions under the excitation wavelengths of 465 nm (blue) and 635 nm (red) using the equivalent circuit model inserted.

While using $TiO_2$ as ETL is dominant in the PSC literature, few articles have explored the low-temperature approaches. The reported plasma-based $TiO_2$ synthesized at low temperatures makes the approach compatible with flexible substrates. To provide a comprehensive overview and performance comparison of $TiO_2$ synthesized at low temperatures as ETL in PSCs, we have compiled a table summarizing critical parameters from relevant studies in the field. **Table 1** summarizes the synthesis method for each study, $TiO_2$ temperature, type of perovskite used, and key photovoltaic parameters, including $V_{OC}$, $J_{SC}$, FF, and PCE.

**Table 1.** Summary of photovoltaic parameters and $TiO_2$ synthesis conditions at low temperatures in previous studies

| Temp (ºC) | Methodology | Voc (V) | Jsc (mA·cm$^{-2}$) | FF (%) | PCE (%) | Ref. | Year |
|---|---|---|---|---|---|---|---|
| RT | spin coating (Ti(IV) isopropoxide) | 1.00 | 20.4 | 71 | 16.3 | [45] | 2017 |
| 70 | Sol-gel + UV ozone | 1.11 | 22.2 | 69 | 17.0 | [37] | 2018 |
| 125 | Spin coating + Steam-Annealing | 1.15 | 22.5 | 73 | 18.9 | [46] | 2018 |
| 150 | Amino functionalization by sol-gel | 1.19 | 23.4 | 75.19 | 21.3 | [35] | 2019 |
| 150 | Chemical Bath | 1.12 | 20.2 | 72 | 14.5 | [47] | 2020 |
| 125 | Spin coating + Steam-Annealing Method | 1.11 | 24.7 | 78 | 21.3 | [36] | 2020 |
| 150 | Annealing inside a domestic MW oven | 1.06 | 21.9 | 70 | 16.4 | [6] | 2022 |
| 150 | Magnetron-Sputtering | 0.96 | 18.6 | 49 | 8.7 | [48] | 2017 |
| 200 | Electron beam + $CO_2$ plasma treatment | 1.04 | 21.4 | 69 | 15.4 | [34] | 2017 |
| 200 | RPVAD-$O_2$* and -Ar* + SPE | 1.03 | 23.6 | 79 | 14.6 | This work | 2025 |

It can be noted that low-temperature approaches below 200ºC have a high potential for revisiting $TiO_2$ as an efficient ETL (see Table 1). However, reported methods are mostly based on wet-chemical approaches such as spin-coating,[6,35–37,45,46] and chemical baths,[47] which are not industrially scalable or environmentally friendly. The reported vacuum/plasma methodology is an energy-efficient process fully compatible with industrial approaches and does not require solvents or toxic chemicals. In the literature survey shown in Table 1, some vacuum and plasma methods have been reported. For example, reactive magnetron sputtering at 150ºC has been used to deposit $TiO_2$ for PSCs with intermediate efficiencies (8.7%).[48] In addition, room temperature electron-beam evaporated $TiO_2$ layers subjected to a $CO_2$ plasma treatment at 200ºC show a high PCE of 15.4%. Moreover, as shown in Table 1, many of the low-temperature $TiO_2$ layers used to develop the most efficient PSCs use post-treatments, which are fully compatible with the layers developed here. For example, amino-functionalization (PCE=21.3%),[35] steam annealing (21.3%),[36,46] microwave annealing (16.4%),[6] or $CO_2$ plasma treatment (15.4%).[34] Thus, through our innovative approach, we have achieved remarkable PSC efficiency (champion PCE of 14.6%), making them very competitive not only with those manufactured using plasma technology but also with wet methods. This accomplishment underscores the effectiveness of our RPAVD method in synthesizing highly efficient perovskite solar cells at lower temperatures while ensuring compatibility and scalability on flexible substrates.

## 4. Conclusions

We have reported the plasma synthesis of porous $TiO_2$ layers using a new methodology that combines remote plasma deposition processes followed by a soft plasma etching treatment performed at low temperatures (below 200ºC) compared to conventional methods. The architecture developed consists first of a Columnar structure (*Columnar+SPE*) with porosity close to 70%, followed by an aerogel-like structure on top (*Columnar-Aerogel*), which increased the total porosity above 85%. Adding the aerogel structure improves the optical and electronic properties, in which its elevated porosity confers the layers with antireflective properties. Although the antireflection is lost when infiltrating the perovskite, the increased porosity of the ETL improves perovskite infiltration, enlarges the surface area, and facilitates charge transport, ultimately contributing to better photovoltaic performance. Moreover, TEM analysis confirms that the external surface of the films is crystalline, predominantly exhibiting rutile and anatase phases. XPS analysis of *Columnar-Aerogel* samples further reveals the absence of reduced Ti species, along with a significant reduction in oxygen vacancies ($O_V$) and hydroxyl ($O_H$) species after plasma etching. These findings indicate that the resulting $TiO_2$ films achieve a crystallinity comparable to high-temperature references, highlighting the effectiveness of this low-temperature approach in producing high-quality, low-defect electron transport layers for perovskite solar cells. Notably, the formation of a crystalline $TiO_2$ phase at such low synthesis temperatures is remarkable, as conventional crystallization typically requires temperatures above 400°C for anatase and 600°C for rutile.

The photovoltaic parameters of PSCs incorporating our plasma-synthesized $TiO_2$ layers reveal significant improvements in Jsc and PCE, reaching competitive values compared to reference cells that use compact and mesoporous $TiO_2$ layers synthesized at high temperatures. Specifically, the *Columnar-Aerogel $C_2$* layers have shown a champion PCE of 14.6% (average PCE=13.0±1.6 %), a remarkable value compared to the optimized reference devices (champion and average PCE of 18.1 and 16.6±0.9 %, respectively) synthesized at high temperatures (450ºC). The initial columnar structure provides robust and conductive support, while the overlaid aerogel layer enhances perovskite infiltration and electrical connectivity. Additionally, impedance spectroscopy studies indicate that our plasma-based layers do not introduce preferential recombination sites, aligning with the XPS and TEM data, demonstrating a high crystallinity with low oxygen vacancy concentrations and no detectable reduced Ti species. The absence of additional surface recombination sites, is reflected in high recombination resistance and constant capacitance over the entire range of analyzed voltages. This behavior is consistent with an improvement in the crystallinity and grain size of the perovskite when the plasma porous layers are used as ETLs. These results indicate that combining the RPAVD+SPE method is a practical approach to synthesizing ETLs and represents a significant advance in PSC technology. Moreover, the reported approach is industrially scalable, environmentally friendly, and offers a viable and efficient alternative for manufacturing flexible and high-performance photovoltaic devices.

**Supporting Information**

Supporting information is available at the end of this document.

**Acknowledgements**

We thank the projects PID2022-143120OB-I00, TED2021-130916B-I00, and PCI2024-153451 funded by MCIN/AEI/10.13039/501100011033 and by "ERDF (FEDER) A way of making Europe, Fondos Nextgeneration EU and Plan de Recuperación, Transformación y Resiliencia".

The authors also want to thank CSIC for its financial support through the Intramural Project PIE 202260I156 and Interdisciplinary Thematic Platform (PTI) Transición Energética Sostenible (PTI-TRANSENER+). Project ANGSTROM was selected in the Joint Transnational Call 2023 of M-ERA.NET 3, which is an EU-funded network of about 49 funding organisations (Horizon 2020 grant agreement No 958174). The project leading to this article has received funding from the EU H2020 program under grant agreement 851929 (ERC Starting Grant 3DScavengers).

# Supporting Information

**Low-Temperature Remote Plasma Synthesis of Highly Porous $TiO_2$ as Electron Transport Layers in Perovskite Solar Cells**

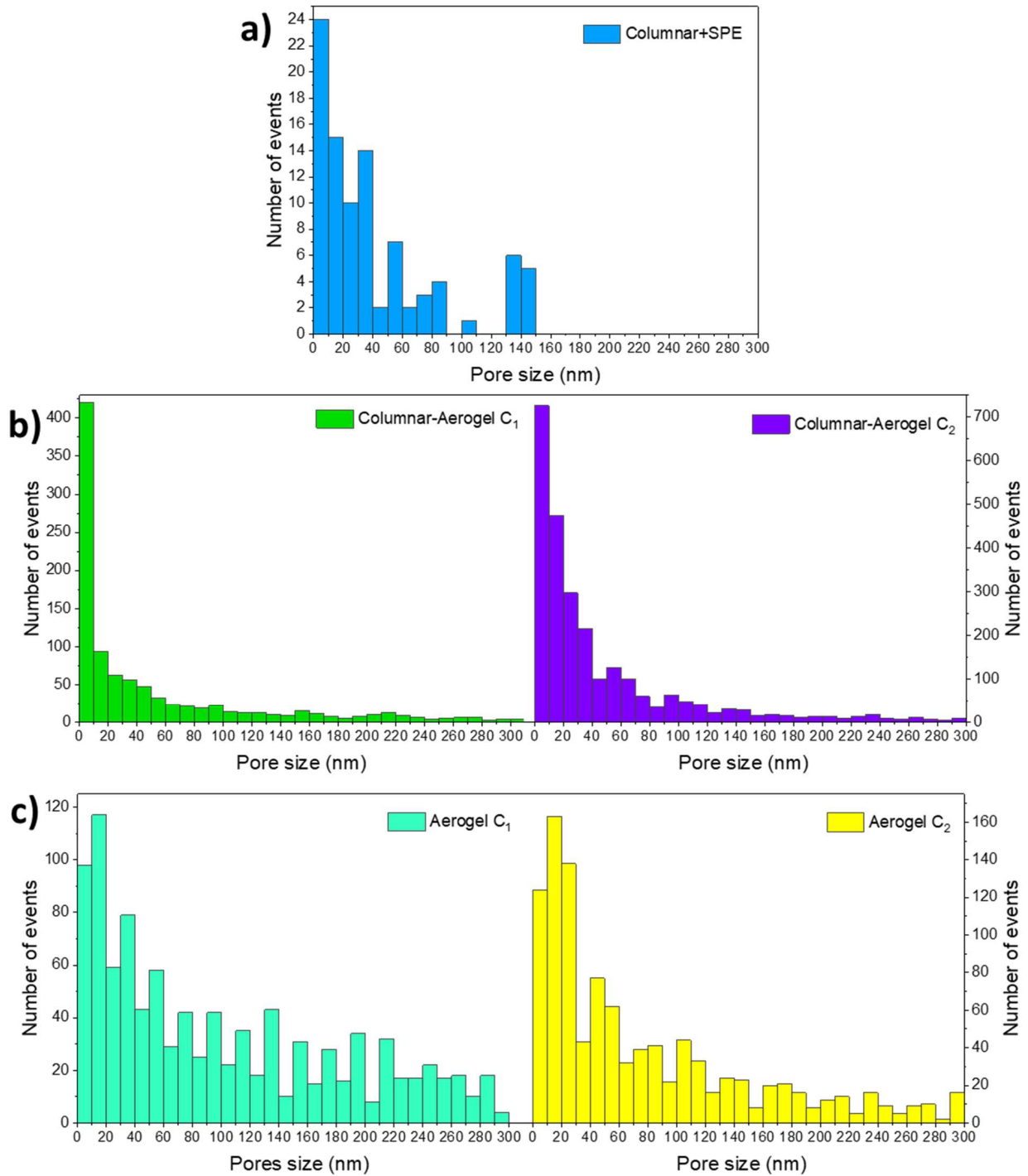

**Figure S1.** Porous structures size distribution of $TiO_2$ samples determined from the analysis of SEM micrographs of a) *Columnar+SPE*, c) *Aerogel-like $C_1$* and *$C_2$* on flat substrates, and b) same thin films on columnar substrates (*Columnar-Aerogel $C_1$* and *$C_2$*).

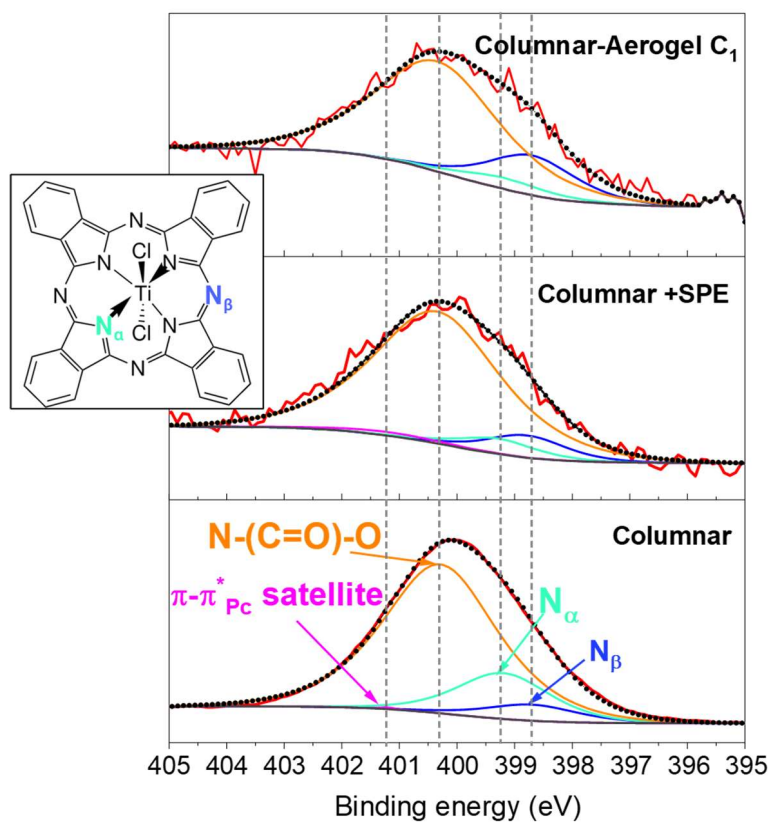

**Figure S2.** Curve fitting of the N1s XPS spectra of the *Columnar* samples before and after SPE treatment, as well as *Columnar-Aerogel C$_1$*. In the spectra, both pyridinic nitrogen (labelled as N$_\beta$) and pyrrolic nitrogen (labelled as N$_\alpha$) at 398.8 eV and 399.3 eV respectively are appreciated.

**Table S1.** Statistical analysis of the photovoltaic parameters of perovskite solar cells derived from current-density versus voltage (J-V) curves measured under 1 sun (AM 1.5G) illumination in reverse scan, using a 0.14 cm² mask. The photovoltaic data for the champion cell is provided in brackets.

|  | $V_{OC}$ (V) | $J_{SC}$ (mA·cm$^{-2}$) | FF (%) | PCE (%) |
|---|---|---|---|---|
| **Reference (mp-TiO$_2$)** | 1.03 (1.02) ± 0.03 | 21.9 (22.8) ± 0.6 | 73 (78) ± 5 | 16.6 (18.1) ± 0.9 |
| **Columnar** | 0.78 (0.96) ± 0.30 | 3.8 (10.0) ± 2.6 | 44 (56) ± 18 | 1.7 (5.3) ± 1.6 |
| **Columnar SPE** | 0.97 (1.00) ± 0.03 | 13.2 (19.0) ± 3.1 | 57 (56) ± 7 | 7.3 (10.6) ± 1.7 |
| **Columnar-Aerogel C$_1$** | 0.95 (0.96) ± 0.05 | 16.5 (18.4) ± 1.9 | 66 (72) ± 8 | 10.4 (12.7) ± 1.5 |
| **Columnar-Aerogel C$_2$** | 1.01 (1.03) ± 0.02 | 20.2 (18.2) ± 2.7 | 64 (78) ± 12 | 13.0 (14.6) ± 1.6 |
| **Aerogel C$_1$** | 0.89 (0.98) ± 0.05 | 2.5 (2.7) ± 0.3 | 38 (44) ± 5 | 0.9 (1.2) ± 0.3 |
| **Aerogel C$_2$** | 0.99 (1.02) ± 0.02 | 6.7 (8.4) ± 1.3 | 53 (56) ± 3 | 3.5 (4.3) ± 0.7 |

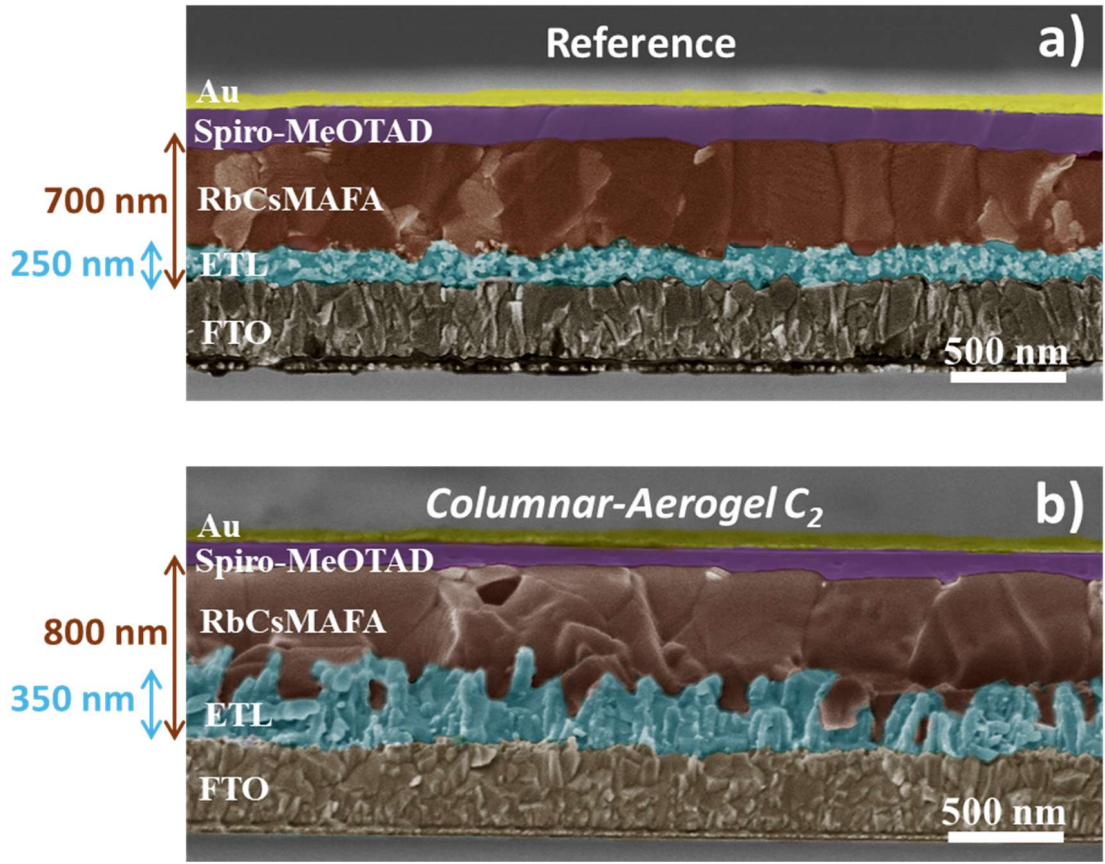

**Figure S3.** Cross-sectional SEM micrographs of complete solar cells, illustrating a comparison between the TiO$_2$ ETL with a) mesoporous TiO$_2$ (used as reference) and b) TiO$_2$ Columnar-Aerogel C$_2$.

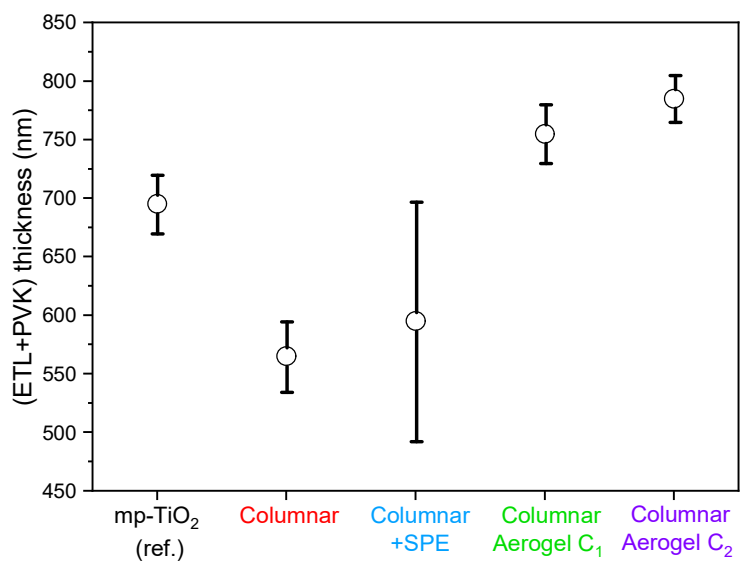

**Figure S4.** Thickness of perovskite deposited on top of plasma-synthesized TiO$_2$ ETLs obtained from cross-section image from SEM.

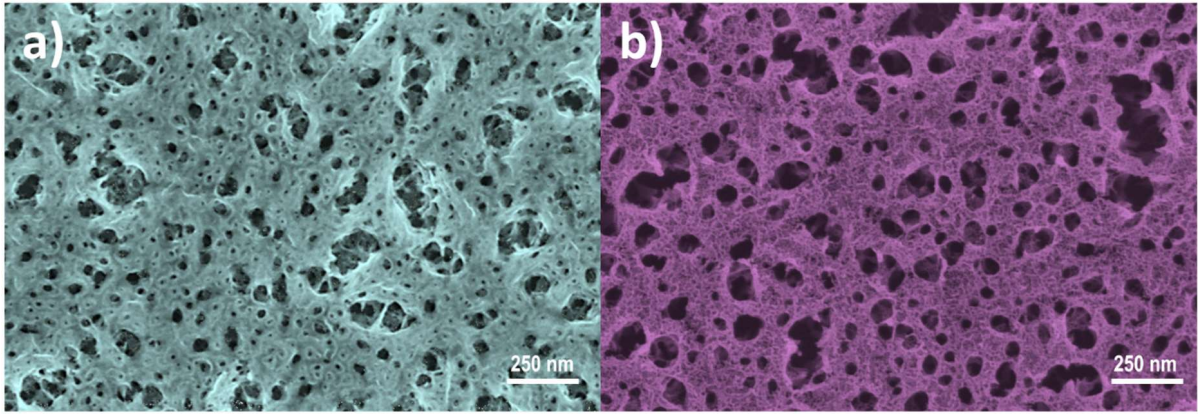

**Figure S5.** Top-view SEM pictures of *Aerogel-$C_2$* (a) and *Columnar+Aerogel $C_2$* (b).

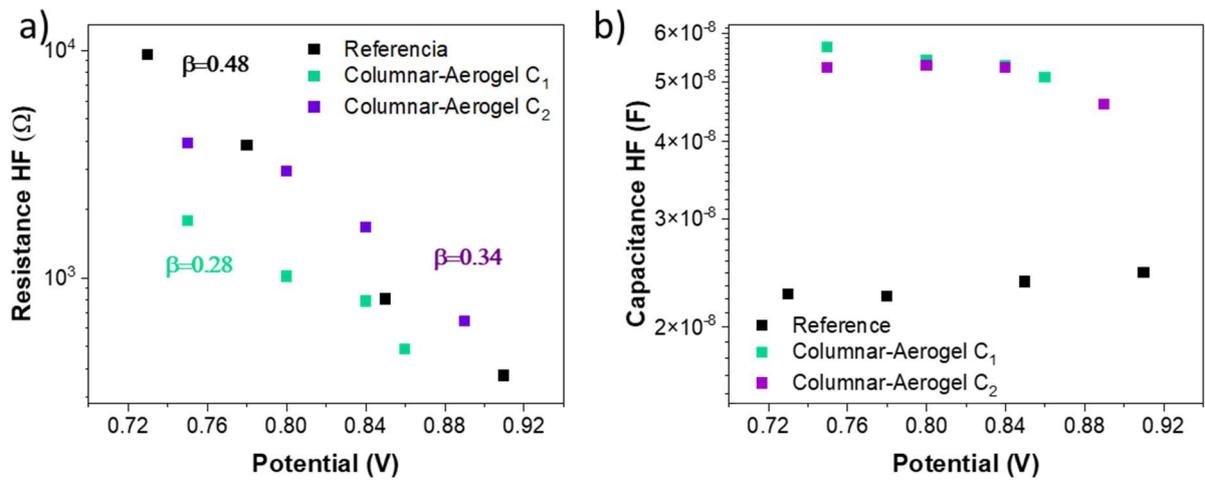

**Figure S6.** High-frequency resistance and capacitance extracted from fittings of the impedance spectra obtained at open-circuit conditions under the excitation wavelengths of 635 nm using the equivalent circuit model inserted in Figure 7 b).

31